\documentclass[journal=jpclcd,manuscript=letter,layout=traditional]{achemso}

\usepackage[version=3]{mhchem} 
\usepackage{subfig}
\usepackage{graphicx}

\author{Niranjan Shivaram}
\email{niranjan@purdue.edu}
\affiliation{Chemical Sciences Division, Lawrence Berkeley National Laboratory, Berkeley, CA 94720, United States}
\alsoaffiliation{Department of Physics and Astronomy, Purdue University, West Lafayette, IN 47907, United States} 
\alsoaffiliation{Purdue Quantum Science and Engineering Institute, Purdue University, West Lafayette, IN 47907, United States}

\author{Richard Thurston}%

\author{Ali Belkacem}
 
 \author{Thorsten Weber}
\affiliation{Chemical Sciences Division, Lawrence Berkeley National Laboratory, Berkeley, CA 94720, United States}

\author{Liang Z. Tan}
\affiliation{Molecular Foundry, Lawrence Berkeley National Laboratory, Berkeley, CA 94720, United States}

\author{Daniel S. Slaughter}
\email{dsslaughter@lbl.gov}
\affiliation{Chemical Sciences Division, Lawrence Berkeley National Laboratory, Berkeley, CA 94720, United States}

\title[two-color four-wave mixing]
  {Interplay between ultrafast electronic and librational dynamics in liquid nitrobenzene probed with two-color four-wave mixing}

\abbreviations{IR,FWM,UV}
\keywords{four-wave mixing, ultrafast, nonlinear spectroscopy, time-dependent polarization response, dephasing, molecular dynamics, electronic and nuclear coupling}

\begin{document}

\begin{tocentry}
\includegraphics[width = 1.0\linewidth]{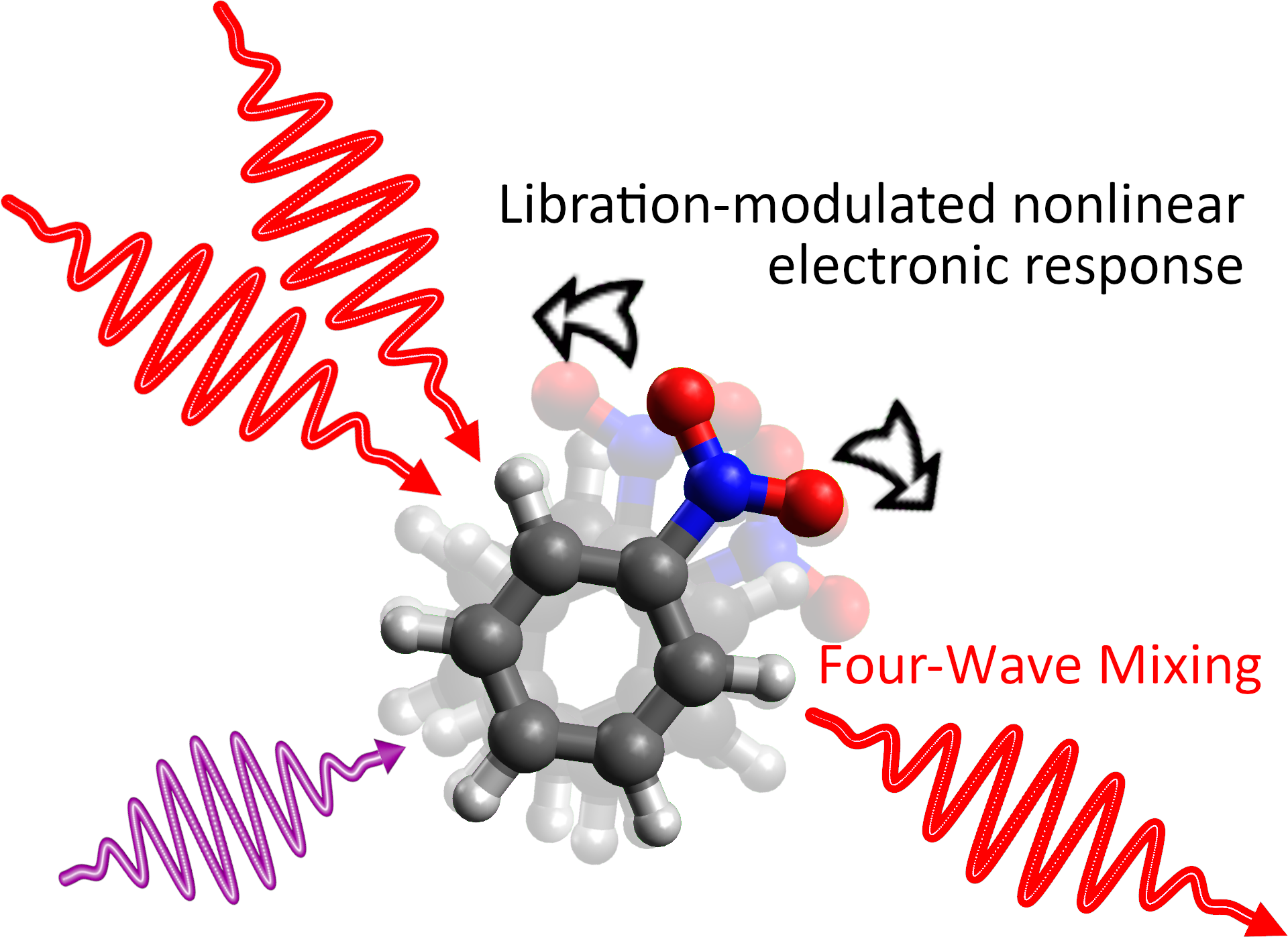}\\
\label{fig:toc}




\end{tocentry}

\begin{abstract}
We present an experimental and theoretical study of the interplay between ultrafast electron dynamics and librational dynamics in liquid nitrobenzene. A femtosecond ultraviolet pulse and two femtosecond near infrared pulses interact with nitrobenzene molecules, generating a four-wave mixing nonlinear signal that is measured in the Optical Kerr Effect geometry. The near infrared nonlinear signal is measured to be non-zero only at negative time delays, corresponding to the near infrared pulses arriving earlier than the ultraviolet pulse. We perform time-dependent Quantum Master Equation calculations, which include a classical libration model, to simulate the experiment. The simulations support the conclusion that the near infrared pulses launch librational motion, while simultaneously creating electronic coherences that result in a libration-modulated electronic nonlinear response. Furthermore, we conclude that the measured nonlinear optical signal corresponds to a non-parametric process that leaves the molecules in an excited electronic state. This work provides new insight into ultrafast nonlinear optical interactions in liquids and is an important step towards probing ultrafast electronic coherences in large molecules in the liquid phase.     
\end{abstract}

Photochemical reactions are driven by the ultrafast coupled motion of electrons and nuclei in molecules, which can further couple with the surrounding environment. Each of these couplings transfer the energy injected by the initial photoexcitation through various electronic decay mechanisms, or nuclear motion such as dissociation, isomerization, vibrational or rotational excitation. Several recent developments in coherent multidimensional nonlinear spectroscopy\cite{cho_coherent_2008,gross_progress_2023,ma_two-dimensional_2023,gaynor_signatures_2017,ramesh_enhancing_2023,han_asynchronous_2025,gajo_two-dimensional_2025,giubertoni_situ_2022} have demonstrated the possibility to reveal the populations of excited electronic states and the coherences between pairs of electronic states in these dynamics. Ultrafast time-resolved nonlinear spectroscopies not only probe the populations and coherences driving the coupled electronic and nuclear dynamics following coherent photoexcitation, but also directly probe the ultrashort electronic decay lifetimes and the timescales of the relevant coupling mechanisms.\cite{scholes_using_2017} 

Recent ultrafast time-resolved nonlinear spectroscopic investigations have revealed coupled electronic and nuclear dynamics in isolated gas-phase molecules\cite{lin_coupled_2021,timmers_disentangling_2019,marroux_multidimensional_2018} and in liquids\cite{runge_nonlinear_2023,ensing_origin_2019,yagasaki_fluctuations_2013}. The analysis of signals in the frequency domain reveals the active degrees of freedom in the excited system. Time-domain analyses and simulations can reveal aperiodic dynamics, such as nonadiabatic transitions,\cite{lavoine_nonadiabatic_2003} or processes occurring within the duration of the laser pulses,\cite{smallwood_analytical_2017,perlik_finite_2017} provided the pulses are well characterized. Measured and calculated coherent polarization interactions have been explored in dense media, where local field effects were found to enhance signals that were emitted for negative delays, i.e., where the probe pulse arrives before the pump pulse(s).\cite{wegener_line_1990, meier_electronic-oscillator_1997} 

For isotropic samples such as gases, liquids, or amorphous films, the lowest order nonlinear optical response is the third-order response, because the second order optical response vanishes in the presence of inversion symmetry,\cite{boyd_non-linear_2008} with few exceptions.\cite{fischer_isotropic_2001} Four-wave mixing (FWM) spectroscopies probe the third order susceptibility of molecules primarily through resonant optical transitions. Many FWM methods apply parametric processes, where the initial and final states of the electronic system are the same. Notable examples of nonparametric schemes, such as coherent (anti-)Stokes Raman scattering, where the initial and final states differ, are well-established and widely applied in vibrational spectroscopic techniques \cite{wright_theoretical_1997}. Recently, nonparametric FWM experiments involving electronic transitions \cite{oneal_electronic_2020-1} have been reported in the literature, opening the possibilities for new spectroscopic schemes \cite{kowalewski_simulating_2017-1} that employ pulses having sub-femtosecond durations, and atomic site specific probes triggering transitions involving inner valence or core orbitals that are localized within a molecule. 

In the liquid phase, an incident ultrashort laser pulse can excite electronic and vibrational transitions and induce electronic polarization, while the brief but intense optical field also typically causes librational motion of the molecules. Librational dynamics in liquids have been extensively studied using nonlinear spectroscopies \cite{ruhman_intramolecular_1987,palese_femtosecond_1994,heisler_polarization-resolved_2008}. However, the influence of librational motion on electronic excitations, and the ultrafast evolution of electronic coherences resulting from such excitations, have not been thoroughly explored. Furthermore, interpretations of nonlinear spectroscopic measurements with quantum mechanical simulations in the time-domain are computationally demanding, particularly for short-wavelength optical pulses, due to the short intervals in real time that are required to accurately simulate the optical pulses. In this paper, we report on the experimental and theoretical investigations of coupled electronic and nuclear libration dynamics in nitrobenzene, by employing ultraviolet and near-infrared pulses to generate FWM in liquid nitrobenzene. The measured signals are interpreted with the aid of time-dependent non-perturbative quantum mechanical simulations to understand the dynamics of the molecule in its environment on femtosecond timescales. 

\begin{figure}
\includegraphics[width = 0.80\linewidth, trim = {1cm 9.5cm 4cm 0.1cm}, clip]{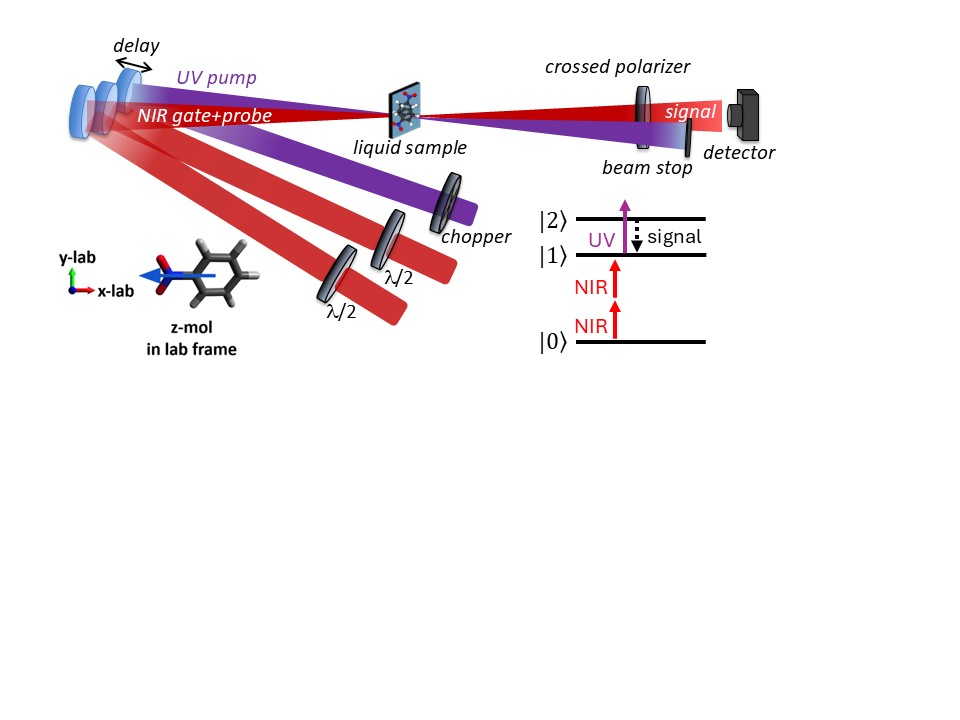}\\
\caption{Schematic diagram of the experimental geometry. A sample of nitrobenzene at room temperature interacts with one UV pump pulse and 2 NIR pulses. Four-wave mixing signals emitted in the NIR Probe direction are detected by a photodiode detector using a lock-in amplifier referenced to an optical chopper acting on the UV pump pulse. The polarization of the NIR pulses are set independently with half-wave plates ($\lambda/2$). Left inset: diagram of the nitrobenzene structure in the selected orientation. Right inset: Energy level diagram of the relevant nonparametric FWM scheme. Arrows representing transitions are not to scale.}
\label{fig:setup-oview}
\end{figure}

The present experiments apply near infrared (NIR) and ultraviolet (UV) pulses generated with a titanium sapphire femtosecond laser system that delivers 780~nm pulses at a repetition rate of 1 kHz, with each pulse having a duration of 50~fs full width at half maximum (FWHM) and an energy of 2~mJ. Beam splitters are used to split the pulse into three separate pulses. One pulse is frequency-tripled using nonlinear crystals in a collinear tripling scheme to convert the 780~nm pulses to 260~nm UV pulses, and the residual fundamental and second harmonic are rejected by dichroic mirrors that transmit 780~nm and 390~nm. The resulting UV pulse duration was measured to be 220~fs FWHM (see Fig.~\ref{fig:XFROG} in the SI). The two NIR pulses are optically delayed relative to the UV pulse, and the polarization of one NIR pulse (gate) is rotated to 135$^\circ$ with respect to the UV pump and NIR probe polarization using a half waveplate (see Fig.~\ref{fig:setup-oview}). The signal emitted along the NIR probe direction is analyzed with a 90$^\circ$ crossed polarizer and a lock-in amplifier, which is referenced to an optical chopper modulating the UV pulse train. The two NIR pulses and the UV pulse are focused non-colinearly to a small volume of the sample with an intensity of 10~GW/cm$^2$ per pulse.

We model the nitrobenzene sample with \textit{ab initio} molecular electronic structure calculations, empirical electronic relaxation and dephasing rates, and classical librational motion.  
To simulate the time evolution of the electronic system of a single nitrobenzene molecule, we first specify the excited state energy levels $\Omega$ and transition dipole moments $\vec{\mu}$  for each pair of electronic states. We then performed theoretical calculations of the emitted signal electric field using Lindblad equation simulations in the time-domain, solving

\begin{equation}\label{eq:lindblad}
\dot{\rho}(t) = -\frac{i}{\hbar}[H(t),\rho(t)] + \mathcal{L}_D \rho(t)
\end{equation}
with the Hamiltonian
\begin{equation}
H(\vec{r},t) = \Omega + \vec{\mu} \cdot (\vec{E}_1(\vec{r},t) + \vec{E}_2(\vec{r},t) + \vec{E}_3(\vec{r},t) )
\end{equation}
containing the applied fields $\vec{E}_1$, $\vec{E}_2$, $\vec{E}_3$ corresponding to the two NIR pulses and one UV pulse. The optical fields are constructed from the measured phase and amplitude of each pulse (see Fig.~\ref{fig:XFROG} in the SI). 

Nuclear dynamics are included by an electric field-dependent force function, currently implemented to describe librational motion of a rigid molecule in a liquid that is interacting with two pulses having a variable femtosecond delay. In this model, nitrobenzene rotates around an axis $x_\textrm{mol}$, here fixed to be orthogonal to $z_\textrm{mol}$ (which is parallel to the C-N bond) and within the benzene plane (see Fig.~\ref{fig:setup-oview}, left inset). The angular position of the molecule is described by a single variable $\theta$, which follows the equation of motion
\begin{equation}
    \ddot{\theta}= - \sum_j\omega_j^2\sin(\theta-\theta_j)- \omega_R^2\sin\theta-\gamma\,\dot{\theta}
\end{equation}
Here, the first term describes the restoring force that aligns the permanent dipole of nitrobenzene to the polarization direction $\theta_j$ of each pulse $j$. The restoring frequencies $\omega_j$ are dynamic, as they should increase with the strength of the instantaneous applied fields $E(t)$. We set $\omega_j^2 = \omega_{lib}^2\frac{E(t)^2}{E_{\rm{max}}^2}$, with $E_{\rm{max}}$ the maximum value of the pulse. We use $\omega_{lib}=113 \,\rm{cm}^{-1}$ for the NIR pulse, corresponding to a fundamental band measured by transient grating spectroscopy\cite{heisler_polarization-resolved_2008}, and $\omega_{lib}=18.7 \,\rm{cm}^{-1}$ for the UV pulse. The second term describes the restoring force that aligns the permanent dipole of nitrobenzene to its equilibrium position in the absence of external fields ($\theta=0$), with restoring frequency $\omega_R=56 \,\rm{cm}^{-1}$. The last term describes damping of librational dynamics with time constant $\gamma^{-1}=0.375 \,\rm{ps}$. The initial orientation and equilibrium position is selected to be representative of an ensemble of randomly-oriented molecules. This orientation was determined by computing the polarization signal in the laboratory frame of reference using time-dependent perturbation theory~\cite{thurston_experimental_2023, thurston_polarization_2025} and identifying a single orientation (see Fig~\ref{fig:setup-oview}, left inset) that represents an orientation-averaged ensemble.\cite{thurston_polarization_2025}. 

To model the electronic structure of nitrobenzene, we performed electronic structure calculations of nitrobenzene in a frozen nuclear geometry, using the Multiconfiguration Self Consistent Field (MCSCF) wavefunction method with the aug-cc-pCVDZ basis set implemented in the Dalton quantum chemistry package \cite{aidas_dalton_2014}. The dipole transitions of nitrobenzene were computed using the response theory method \cite{jonsson_response_1996} and the double zeta augmented Dunning correlation consistent basis, with an active space consisting of 14 active electrons that are distributed among 11 orbitals. While the equilibrium geometry of nitrobenzene is a planar C\textsubscript{2v} symmetric molecule, prior gas phase measurements\cite{domenicano_molecular_1990} suggest that the average geometry of nitrobenzene has a $13^\circ$ dihedral angle between the NO\textsubscript{2} functional group and the plane of the benzene ring. As such, the present electronic structure calculations use this twisted geometry of nitrobenzene with C\textsubscript{1} symmetry. These calculations were used to parameterize the transition dipole moments $\vec{\mu}$ of a model Hamiltonian consisting of the ground state $S_0$, and energetically low-lying excited states $S_1$ and $S_2$ all of which have no symmetry due to the lack of symmetry of nitrobenzene in the twisted average geometry. The electronic structure parameters are summarized in the SI, Table~\ref{table:mus}. 

The electronic structure model is then applied in a time-dependent calculation of the nonlinear polarization response to three optical pulses in Liouville space for electronic decay and dephasing rates specified empirically at 100 - 750~fs (see Table~\ref{table:taus} in the SI) in the Lindbladian $\mathcal{L}_D$. The Lindblad equation is numerically solved using the Euler method with fixed time step of 0.1~fs, using the UTPS simulation package.\cite{tan_utps_nodate} This approach  allows the time-dependence of both electronic populations and electronic coherences to be computed. We found this low computational cost time-dependent approach to be adequate for modeling the present femtosecond time-resolved  FWM experiments with two NIR pulses and one UV pulse interacting with liquid nitrobenzene. 

The result of solving Eq.~\ref{eq:lindblad} is the time domain polarization $\vec{P}(\vec{r},t) = \mathrm{Tr}[\vec{\mu} \rho(\vec{r},t)]$. To extract the third-order nonlinear signal electric field, we impose phase matching conditions by selecting only wavevectors parallel to a propagation direction $\vec{k}$. 
\begin{equation}\label{eq:phase-scanning}
\vec{P}_{\mathrm{sig}}(\omega,\vec{k}) =\int \!\! dt \,\, e^{i\omega t} \int \!\! d^3r \,\, e^{-i\vec{k}\cdot\vec{r}} \vec{P}(\vec{r},t)   
\end{equation}
with $\vec{k}=n_1\vec{k_1}+n_2\vec{k_2}+n_3\vec{k_3}$ being the signal wavevector corresponding to the phase matching conditions for given integers $n_1,n_2,n_3$. The polarization response is computed at each time-step by scanning the relative phases of the optical fields in real space ($\vec{E}_j(\vec{r},t)\propto e^{i(\vec{k}_j\cdot\vec{r}-\omega_j t)}$) and performing the Fourier transforms in Eq.~\ref{eq:phase-scanning} to extract the specified phase matching and frequency components.

\begin{figure}
\centering
\includegraphics[width = 0.6\linewidth, trim = {0cm 6.0cm 20.0cm 0.1cm}, clip]{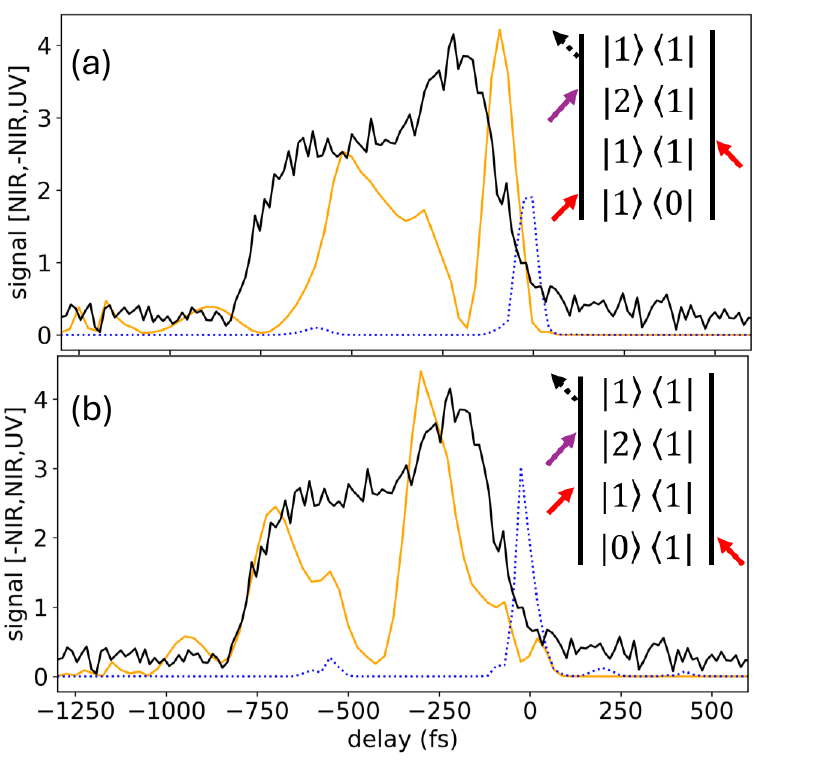}
\caption{Panels (a-b): Measured (solid black curves) and simulated four-wave mixing signals for two 3-level electronic models of nitrobenzene with librational dynamics. Each panel shows the delay dependence of the signal produced by the phase matching condition (see Eq.~\ref{eq:phase-scanning}) that is indicated on the vertical axis. The blue-dotted curves show the signal for the semi-empirical 3-level electronic model, where the energy of the first excited state was calculated to be E$_{S1}$~=~4.17~eV above the ground state S$_0$. The gold curves show the results from the same model but with a modified energy E$_{S1}$~=~3.22~eV. Each simulated signal is rescaled for comparison with the measured signal (in arbitrary units). At negative delays, the time-overlapped NIR pulses arrive before the UV pulse. Insets: Double-sided Feynman diagrams representing selected nonparametric FWM interactions contributing to the simulated signals in panels (a) and (b).}
\label{fig:sim-results}
\end{figure}

The measured FWM signal is maximum for gate-probe time-overlap and decays symmetrically for non-zero gate-probe delays (see Fig.~\ref{fig:signal-2D} in the SI). For the results reported here, we have fixed the gate-probe delay at 0~fs. 
The UV-NIR delay-dependent FWM signal is presented in Fig.~\ref{fig:sim-results} (black curves), exhibiting a double-peak structure for pump-gate delays between -900 and 100~fs. In the convention adopted here, negative delays correspond to the NIR pulses arriving before the UV pump pulse. The FWM signal increases from -800~fs, exhibiting a local maximum at -600~fs. A shallow local minimum between -500~fs and -400~fs is followed by a second local maximum at -200~fs, which is about 32\% higher than the peak at -600~fs. We note that each peak has a width considerably broader than the UV-NIR cross correlation of 220~fs [see Fig.~\ref{fig:XFROG}(a)], while the total FWM signal spans a much longer range of about 950~fs in the UV-NIR delay.  

In Fig.~\ref{fig:sim-results}(a and b) we show the results of our time-dependent calculations (gold solid and blue-dotted curves) of the FWM signals for the selected phase matching conditions indicated on the vertical axes. The computed signal is isolated by phase scanning (Eq.~\ref{eq:phase-scanning}) and frequency filtering to ensure that the signal is consistent with the measured FWM signals (black curves). For each of the phase-matching conditions, we observe the emergence of a double-peak structure in the simulated signal. The double-sided Feynman diagram corresponding to each phase-matching condition, for signals emitted at negative delays, are illustrated as the insets in Fig.~\ref{fig:sim-results} (a and b). 

We assign the simulated signals in Fig.~\ref{fig:sim-results}(a) (gold and blue-dotted curves) to the interaction scheme represented by the corresponding double-sided Feynman diagram (inset of panel (a)). Similarly, the simulated signals of Fig.~\ref{fig:sim-results} (b) are assigned to the inset diagram of panel (b). These two interactions differ only in the time ordering of the NIR interactions on the $|S_{j}\rangle$ and $|S_{i}\rangle$ states. Each calculation follows the approach described above, employing a simplified 3-level electronic structure model and classical librational nuclear motion, in order to investigate the dynamics underlying the measured FWM signals. We surveyed several more complex electronic structure models, consisting of up to 5 states, which we found all produce similar results to the present simple model, but with a higher computational cost. More importantly, we found that the time dependent signals change dramatically with the energy of the first excited state S$_1$ between 4.17~eV (blue-dotted curves in Fig.~\ref{fig:sim-results}) and 3.22~eV (gold curves). The lower S$_1$ potential energy is selected to explore the possible effect of nonadiabatic relaxation of the lowest singlet excited state to the ground electronic state S$_0$ by internal conversion or to the nearly-degenerate manifold of triplet states by inter-system crossing \cite{thurston_ultrafast_2020}. We note the molecular geometry near the conical intersection between S$_1$ and S$_0$\cite{giussani_insights_2017}, where the C-N bond is compressed from 1.48~\AA~to 1.25~\AA, the N-O bonds are both stretched from 1.24~\AA~to 1.32~\AA, and the O-N-O angle closes from 125.0$^\circ$ to 95.8$^\circ$. The potential energy in that region of the potential energy landscape was calculated to be 3.20~eV,\cite{giussani_insights_2017} and the S$_1$ minimum was determined at 2.66~eV.

\begin{figure}
\includegraphics[width = 0.6\columnwidth, trim = {12cm 1.0cm 10cm 0.1cm}, clip]{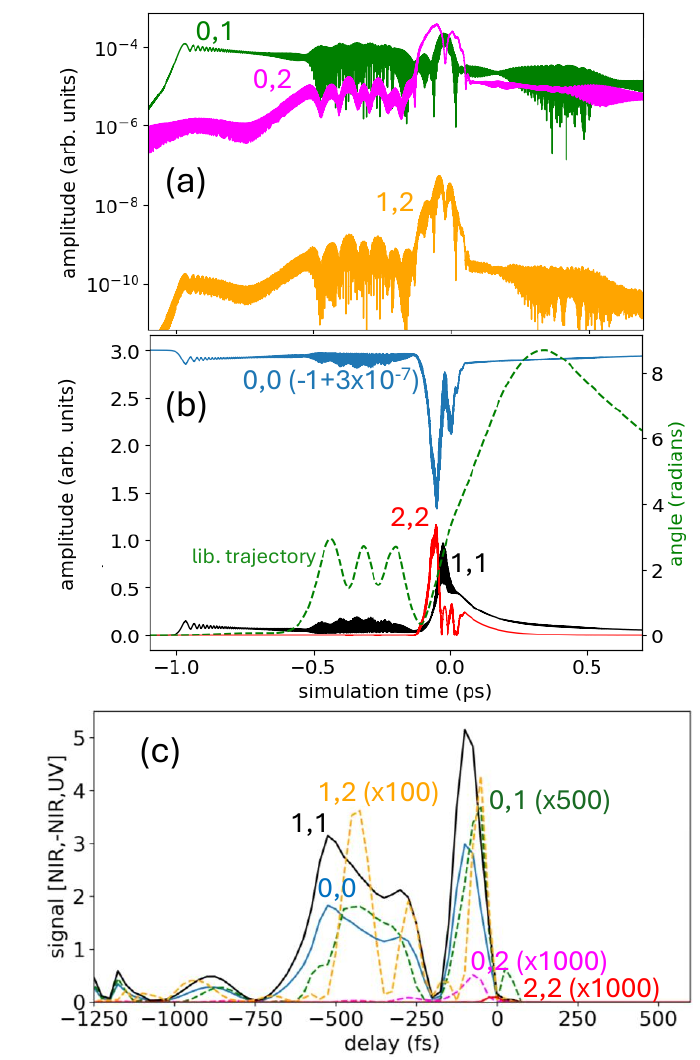}
\caption{Simulated real-time amplitudes of (a) the coherences $|S_{i}\rangle \langle S_{j}|$ and (b) populations $|S_{i}\rangle \langle S_{i}|$ as a function of simulation time for the 3-level electronic structure model having a reduced energy for the first excited state, E$_{S1}$ = 3.22~eV, but otherwise identical to the semi-empirical electronic structure calculation, producing the signals shown as the gold solid curves in Fig.~\ref{fig:sim-results}. In panel (b), we also show the librational trajectory (green dashed curve, right axis). The maxima of the NIR pulses are at -0.125~ps, and the UV pulse maximum is at 0~ps. (c) The contributions as a function of delay of each density matrix element $|S_{i}\rangle \langle S_{j}|$ to the simulated signal with phase matching condition [NIR,-NIR,UV] (gold curve in Fig.~\ref{fig:sim-results}(a). At negative delays, the NIR pulses arrive before the UV pulse.} 
\label{fig:comps-and-diagrams}
\label{fig:signal-comps}
\end{figure}

The simulation-time dependent amplitudes of each density matrix element are extracted from the simulations and shown in Fig.~\ref{fig:signal-comps}(a and b); they correspond to the calculated signals of Fig.~\ref{fig:sim-results} (gold solid curves). The coherence amplitudes (off-diagonal density matrix elements) are displayed on a log scale in Fig.~\ref{fig:signal-comps}(a), and the population amplitudes (diagonal terms) are displayed on a linear scale in Fig.~\ref{fig:signal-comps}(b). Fig.~\ref{fig:signal-comps}(c) shows the delay-dependence of the coherences and populations. The $|S_{1}\rangle \langle S_{2}|$ coherence (gold dashed curve in Fig.~\ref{fig:signal-comps}(c)) provides the strongest coherence contribution to the simulated signal of Fig.~\ref{fig:sim-results}(a) (gold solid curve), and the corresponding interaction diagram (Fig.~\ref{fig:sim-results}(a) inset). The populations $|S_{1}\rangle \langle S_{1}|$  and $|S_{0}\rangle \langle S_{0}|$ black and blue solid curves, respectively, in Fig.~\ref{fig:signal-comps}(c) are the strongest contributors to the same signal. A similar trend is found for the signal displayed in Fig.~\ref{fig:sim-results}(b) (gold solid curve) that results from the other relevant interaction [diagram inset in Fig.~\ref{fig:sim-results}(b)].

\begin{figure}
\includegraphics[width = 0.95\columnwidth]{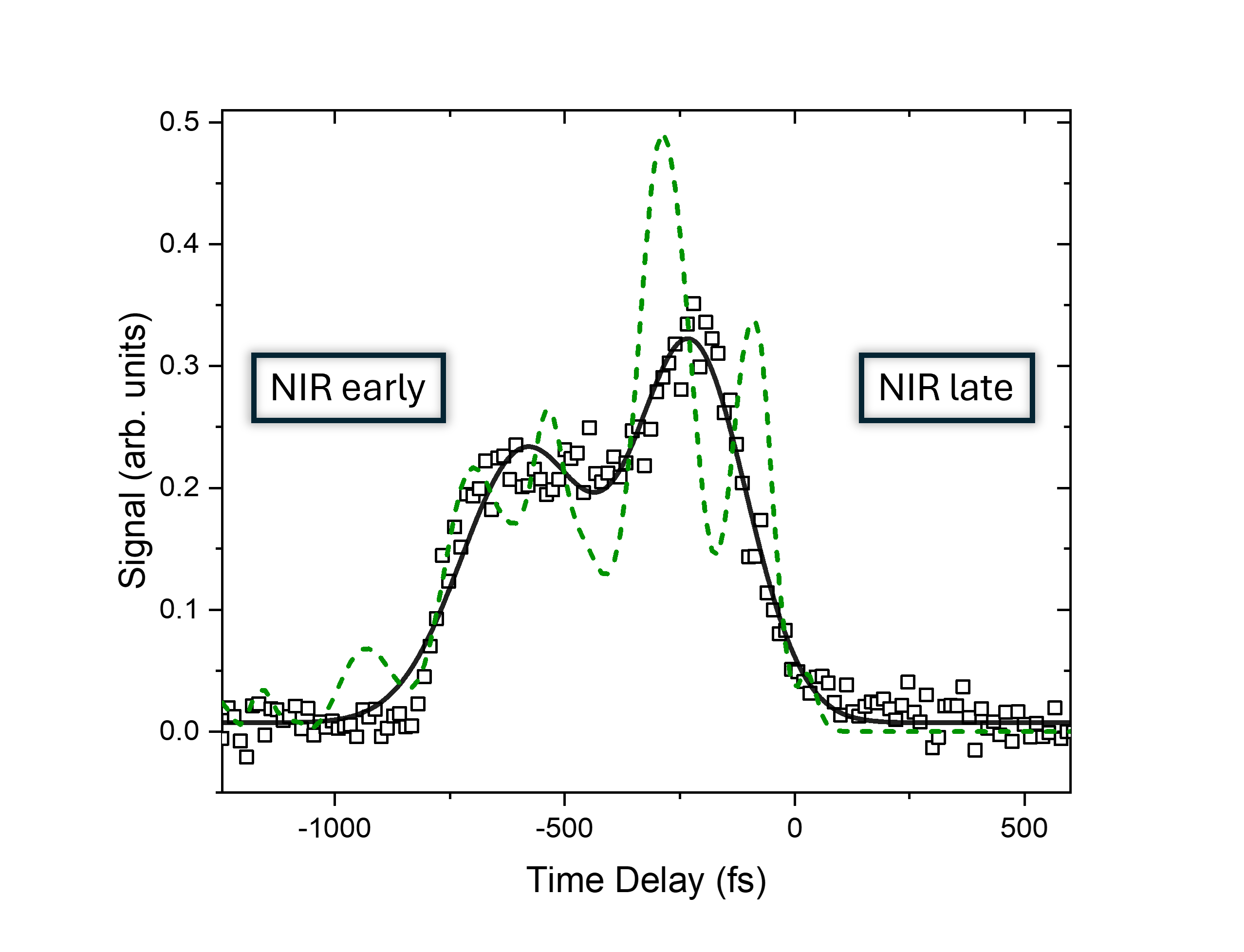}\\
\caption{Measured polarization signal (black square symbols) as a function of delay between the NIR and UV pulses. Contributions from only the NIR pulse pair are excluded from this three-pulse signal by lock-in homodyne detection referenced to a chopper acting on the UV pump pulse. The signal generated with UV and gate pulses (when the probe is blocked) is subtracted from the three pulse signal. The solid black curve is a fit using two Gaussian functions to serve as a guide to the eye. At negative delays, the NIR pulse arrives before the UV pulse. The dashed green curve is a linear combination of the three simulated cases shown in Fig. 2. The simulation agrees well with experimental data.}
\label{fig:exp-results}
\end{figure}

In Fig.~\ref{fig:exp-results}, the observed FWM signals occur mainly for delays where the UV pulse interacts with the sample after the NIR pulse (i.e., delays $<$~0~fs). The NIR pulse, formed by the overlap of the gate and probe pulses, interacts with nitrobenzene in the ground electronic state by 2-photon excitation, establishing a coherent electronic wavepacket that populates the lowest few excited electronic states \cite{giussani_insights_2017,giussani_photorelease_2022,saalbach_ultraviolet_2021,hegazy_tracking_2024}. 
This specific scenario of two NIR interactions and one UV interaction is modeled to produce the simulated signals in Fig.~\ref{fig:sim-results}, and the individual contributions from each population and coherence in Fig.~\ref{fig:signal-comps}. 

As the experimental geometry confines the measured signal to be along the probe direction, phase matching restricts the relevant interaction paths to those that produce the signal in the same direction. The most relevant interaction pathways are summarized in diagrams of Fig.~\ref{fig:sim-results}(a-b). The interactions are all nonparametric, leaving the molecule with an excited state population (See Fig.~\ref{fig:setup-oview}, left inset). The population $|S_{1}\rangle \langle S_{1}|$ is expected to subsequently undergo nonradiative decay via coupled electronic and nuclear motion on $\approx$100~fs to few-ps timescales. Nonetheless, the contributions of librational nuclear dynamics to the measured signals can be described by the present simplified model. \
The observed signals are the result of the electronic polarization of the model system following the nuclear motion trajectories associated with libration. In the experiment, we expect not only librational dynamics similar to the model, but also nuclear motion associated with relaxation of the relevant excited electronic states to contribute to the measured signal. As previously established for nitrobenzene \cite{giussani_insights_2017}, following the excitation of an electronic population in the first singlet state S$_1$, the nitro functional group moves towards O-N-O angle closing and C-N bond shortening, decreasing the potential energy of S$_1$ to approach an inter-system crossing (ISC) between S$_1$ and the second excited triplet state T$_2$. Part of the nuclear wave-packet continues on the potential energy surface of the T$_2$ state, and the remaining portion of the wavepacket continues along the S$_1$ surface, primarily with further C-N shortening and O-N-O angle closing to 94.77$^{\circ}$, eventually reaching the conical intersection (CI) between S$_1$ and S$_0$. Including such highly-dimensional nuclear motion in the present computational model is currently computationally intractable and beyond the scope of the present study. Nonetheless, we expect such dynamics to influence the relevant transition dipole moments, just as the librational dynamics and the transition energies do, which are static in the present model. 

Nonparametric FWM necessarily leaves a population in one or more excited state(s), as illustrated by the simulated density matrix elements in Fig.~\ref{fig:signal-comps}(b). In the present experiment, such FWM signals are expected to transfer momentum to the molecule, relaxing the strict angular dependence of the signal for the phase matching conditions resulting from the interaction diagrams of Fig.~\ref{fig:sim-results}. In Fig.~\ref{fig:exp-results}, we compare the measured FWM signal with a linear combination of the relevant simulated signals [gold solid curves in Fig.~\ref{fig:sim-results}(a-b)], displayed as a green dashed curve. The resulting average of the two simulations is the result of a linear least squares fit, which assigns 41\% and 59\% contributions from the signal and interactions in Fig.~\ref{fig:sim-results} (a) and (b), repsectively. We find the resulting average simulated signal to provide the nearest agreement with the experiment, within the parameter space provided by the 3-level electronic structure model, libration, and decay parameters. 

With a combination of ultrafast time-resolved FWM spectroscopy experiment and theory, we have demonstrated the role of electronic decay, electronic dephasing, and librational motion on FWM signals. FWM of two NIR pulses and one UV pulse in a pure molecular liquid was found to produce femtosecond UV-NIR delay dependence signals that are highly sensitive to both electronic and nuclear dynamics. Ultrafast time-resolved FWM signals, measured in the direction of the NIR probe pulse by polarization spectroscopy, are well-reproduced by time-dependent theory if both librational nuclear motion and electronic relaxation of the first and second singlet excited electronic states are included in the theoretical model. This work showcases the interplay between electronic excitations and librational motion of molecules due to a constrained liquid phase environment, which results in a libration-modulated electronic nonlinear optical response. The NIR pulses participating in the FWM interaction play a dual role; they launch librational motion and create electronic excitations simultaneously. Additionally, our calculations support the conclusion that the measured signals correspond to an electronic non-parametric FWM process, a process rarely observed previously. This work establishes a foundation for future experiments that, for example, employ shorter optical wavelengths to access electronic states in the ionization continuum and enable optical transitions involving inner valence orbitals in molecules to provide atomic site-specificity, especially in the liquid phase. Although the theoretical models presented in this work do not include intramolecular nuclear dynamics primarily to constrain the computational costs, the methodology presented here is extendable to include real-time molecular quantum dynamics. 

\begin{acknowledgement}
Work at LBNL was supported by the US Department of Energy (DOE), Office of Science (Sc), Division of Chemical Sciences Geosciences and Biosciences (CSGB) of the Office of Basic Energy Sciences (BES) under
Award No. DE-AC02-05CH11231. Data analysis was provided by a user project at the Molecular Foundry, supported by the DOE, Sc, BES under Contract No. DE-AC02-05CH11231. This research used the Lawrencium computational cluster resource provided by the IT Division at the Lawrence Berkeley National Laboratory. NS acknowledges support from the DOE, Sc, BES, CSGB under Award No. DE-SC0024234.
\end{acknowledgement}
\clearpage
\renewcommand{\theequation}{S\arabic{equation}}
\renewcommand{\thetable}{S\arabic{table}}
\renewcommand{\thefigure}{S\arabic{figure}}
\setcounter{equation}{0}
\setcounter{table}{0}
\setcounter{figure}{0}

\begin{suppinfo}
\begin{figure}
\includegraphics[width = 0.95\columnwidth, trim = {1cm 1cm 1cm 1.15cm}, clip]{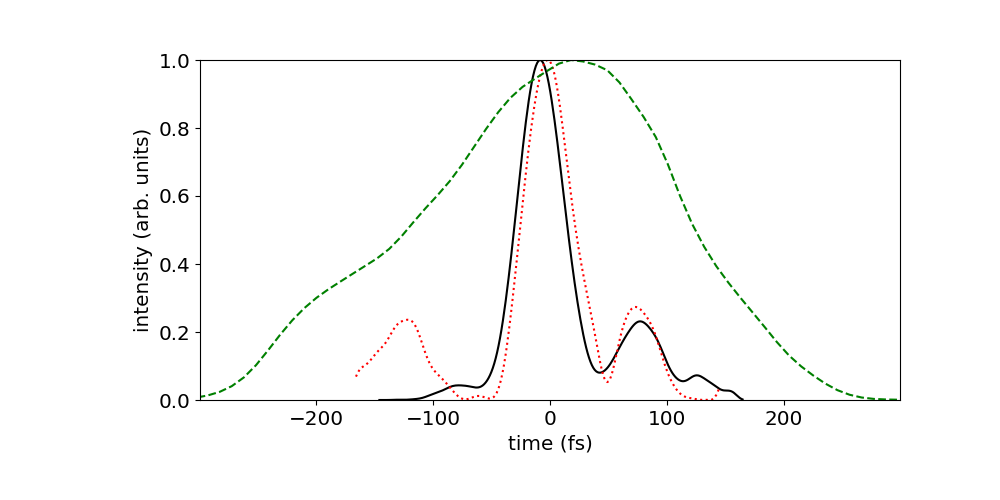}
\caption{Intensities of the NIR probe pulse (black solid curve), NIR gate pulse (red dotted curve) and UV pulse (green dashed curve) as a function of time, measured by cross-correlation frequency-resolved optical gating\cite{linden_amplitude_1999}. 
}
\label{fig:XFROG}
\end{figure}

\begin{figure}
\includegraphics[width = 0.95\columnwidth]{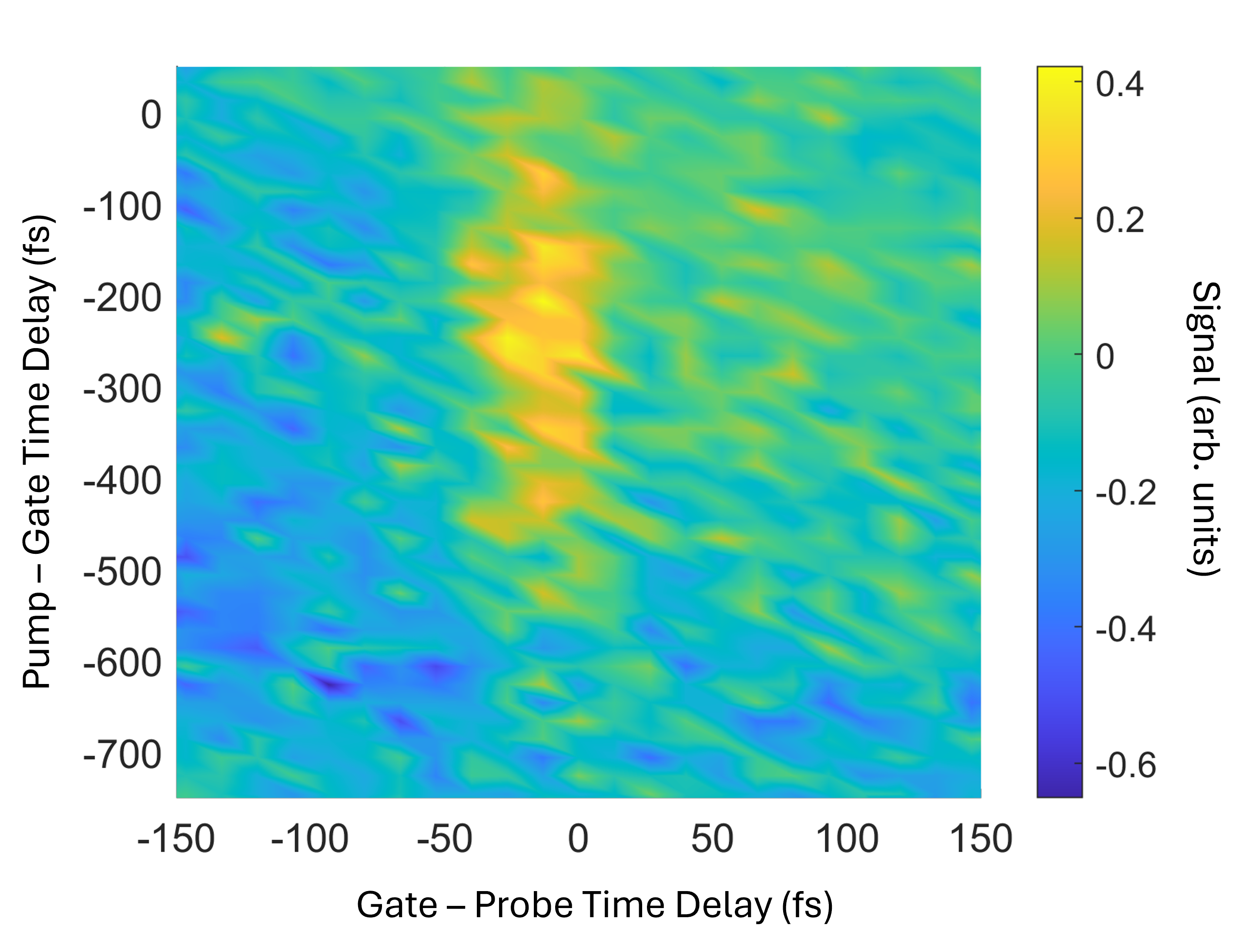}\\
\caption{Measured four-wave mixing signal as a function of delay between the gate and probe pulses (horizontal axis) and the pump and gate pulses (vertical axis). The gate-probe delay is set to zero in the experimental data shown in Fig.~\ref{fig:exp-results}. The data here has lower statistics compared to Fig.~\ref{fig:exp-results}, due to the significantly longer time required to perform a two-dimensional time-delay scan.}
\label{fig:signal-2D}
\end{figure}

\begin{figure}
\includegraphics[width = 0.95\columnwidth, trim = {12cm 6cm 8cm 0.1cm}, clip]{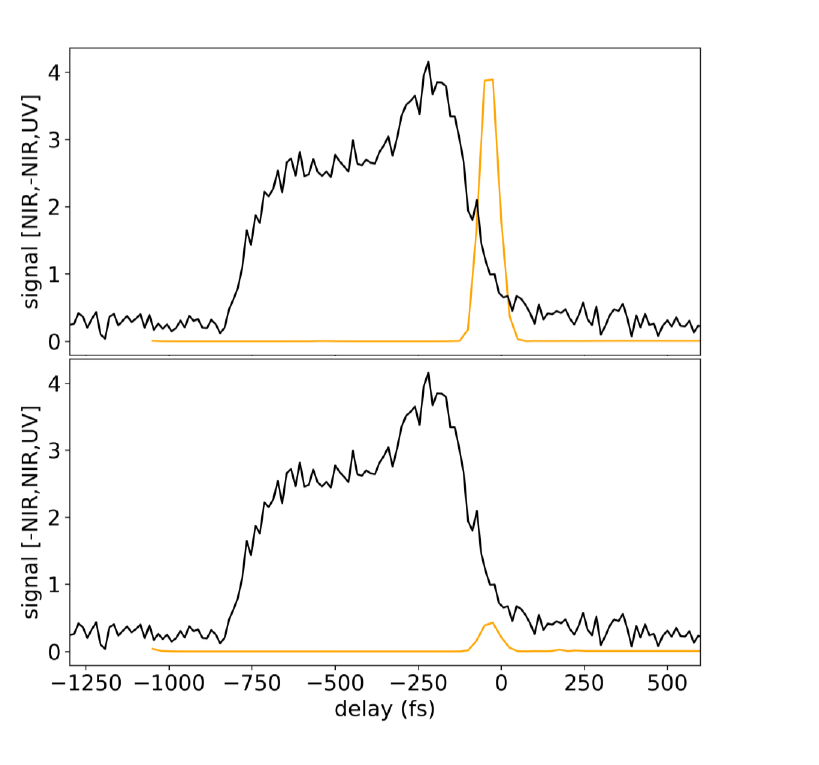}\\
\caption{Simulated four-wave mixing signals (gold solid curve) for a 3-level fixed-nuclei electronic model of nitrobenzene, showing negligible time-dependent signal without nuclear motion, compared to the experimental result (black curve). For this comparison, the simulated signals are multiplied by 10$^9$ relative to the simulations of Fig~\ref{fig:sim-results}(a) and (c).
}
\label{fig:fixed-nuclei}
\end{figure}

\begin{table}
\centering
\begin{tabular}{ l l l l}
\hline
j = & 0 & 1 & 2  \\
\hline
\hline
E$_j$ & 0 & (4.17) 3.32 & 4.56\\
$\mu_{0j}$ & [0, 0, 1.72] & [2.68e-10, -9.03e-11, 0.074] & [0.010, -0.033, -5.96e-10]\\
$\mu_{1j}$ & [2.68e-10, -9.03e-11, 0.074] & [4.90e-05, 2.05e-06, -2.25] & [0.58, 0.14, 5.28e-06]\\
$\mu_{2j}$ & [0.010, -0.033, -5.96e-10] & [0.58, 0.14, 5.28e-06] & [8.01e-05, 6.35e-06, -2.18]\\
\hline
\end{tabular}
\caption{Energies E$_j$ (eV) and transition dipole moments $\vec{\mu_{ij}}=$[$\mu_{x,ij}$, $\mu_{y,ij}$, $\mu_{z,ij}$] (atomic units) coupling the electronic states $|S_{i}\rangle$ and $|S_{j}\rangle$ for the 3-level models employed in the present work. The first electronic structure model (blue-dotted curves in Fig.~\ref{fig:sim-results}), and the second model (gold solid curves in Fig.~\ref{fig:sim-results}) has  E$_1$ = 3.32~eV.}
\label{table:mus}
\end{table}

\begin{table}
\centering
\begin{tabular}{l l l l }
\hline
j = & 0 & 1 & 2  \\
\hline
\hline
$\tau_{0j}$ & - & - & -\\
$\tau_{1j}$ & [0.75, 0.75] & - & -\\
$\tau_{2j}$ & [0.75, 0.75] & [0.10, 0.30] & -\\
\hline
\end{tabular}
\caption{Electronic relaxation and dephasing times [$\tau_{r,ij}$,$\tau_{d,ij}$] (ps) for the 3-level models employed in the present work.}
\label{table:taus}
\end{table}
\end{suppinfo}

\clearpage

\bibliography{UVNIRFWMbib}

\end{document}